\title{A Quantum Mechanical Travelling Salesman}
\author
{
  \textsc{Ravindra Rao}\\
  108 West 41st Street,\\
  Austin, Texas 78751-4610, USA\\
}
\documentclass{article}

\usepackage[margin=2.5cm]{geometry}
\usepackage{amsmath}
\usepackage{bbold}
\usepackage{float,subfig}
\usepackage{pstricks,pst-node,pst-tree}
\psset{unit=1cm}

\def\qg{$\mathcal{G}$ }

\def\ket#1{\mathinner{|{#1}\rangle}}

\def\gstate#1{{\ket {g({#1})}}}

\def\stepop#1{\hat S(#1)}

\def\up#1{\hat {#1}^*}
\def\down#1{\hat {#1}}

\def\upv#1{v_{#1}\!=\!1}
\def\downv#1{v_{#1}\!=\!0}

\def\upa#1#2{a_{#1#2}\!=\!1}
\def\downa#1#2{a_{#1#2}\!=\!0}

\floatstyle{boxed}
\restylefloat{figure}

\begin{document}
\date{}
\maketitle

\begin{abstract}
A quantum simulation of a travelling salesman is described.  A vector
space for a graph is defined together with a sequence of operators
which transform a special initial state into a superposition states
representing Hamiltonian tours.  The quantum amplitude for any tour is
a function of the classical cost of travelling along the edges in that
tour.  Tours with the largest quantum amplitude may be different than
those with the smallest classically-computed cost.
\end{abstract}

\section{Introduction}

In the problem of the travelling salesman we are given a graph with a
fixed number of vertices.  Every vertex is connected to every other
vertex by one edge.  A vertex is not connected to itself.  Every edge
has a cost which is a positive real number.  A Hamiltonian tour of
such a graph is a path which begins at a special vertex, chosen
arbitrarily but then fixed.  Starting from this vertex, a sequence of
steps is taken in each of which a previously unvisited vertex is
visited by travelling along the edge connecting the two.  The cost of
travelling along that edge is accumulated.  The classical cost of a
tour is the sum of the costs associated with the edges traversed
during the tour.  The tour ends when all vertices have been visited
exactly once at which point one more step is taken to return to the
starting vertex.  For a graph with $N$ vertices, there are $(N-1)!$
tours that begin and end at the starting vertex.  The problem is to
find the tour with the minimum cost.  This is the simplest version of
the travelling salesman problem \cite{tspbook:1985}.

In section 2 I will describe a quantum simulation of this problem.
The simulation occurs in a vector space for a graph.  This vector
space is sufficient to describe Hamilitonian tours which are special
basis states.  The simulation begins by placing the graph in a basis
state corresponding to a choice of the starting vertex.  A sequence of
operators will transform this initial state to a final state which is
a superposition of only the Hamiltonian tours.  In this final state
all tours have the same quantum amplitude.  There are three goals for
the description in section 2.  The first is to describe the vector
space for a graph.  The second is to show how this vector space is
capable of describing a Hamiltonian tour.  The last goal is to
construct operators which perform the transformations which may be
interpreted as carrying out all Hamiltonian tours.

The simulation is improved in section 3 by taking into account the
cost of travelling along the edges.  The operators described in
section 2 are modified so that the quantum amplitude of a path is
built up as the tour proceeds.  The amplitude is now not a constant,
but a function of the cost along the edges.  Hamiltonian tours with
smaller classical costs have larger quantum amplitudes.

The dynamical system described is not a quantum algorithm that can
execute on a universal quantum Turing machine described by Deutsch
\cite{Deutsch:1985}; it is a simulation of a special system by a
quantum mechanical one.  Such sytems
\cite{Benioff:1982,Feynman:1985,Lloyd:1996} have long been in
existence.  The simulation requires a quantum mechanical computer
capable of executing it.  The model of a computer most suitable has
been described by Feynman \cite{Feynman:1985} based on work of Benioff
\cite{Benioff:1982} and others\footnote{See references in
  \cite{Benioff:1982,Feynman:1985,Lloyd:1996}.}.

Feynman describes a universal quantum computer which is a one
dimensional lattice of spin-half systems with nearest neighbour
interactions coupled to a system of logic gates.  The state of the
complete system consists of the state of the spin lattice and the
state of the registers on which the gates act.  The Hamiltonian for
the system defines the dynamics of the lattice of spins as well as
that of the gates.  Nothing in Feynman's description requires the
system to be a set of logic gates; they can be more general
\cite{Lloyd:1996}.  In the simulation of the travelling salesman
presented here, the state space of the registers are replaced by the
state space of a graph while the operators which act on the registers
are replaced by the operators for carrying out a Hamiltonian tour.
The system coupled to the lattice of spins is not a set of logic gates
but one which operates on a graph.

\section{Simulating Hamiltonian tours}

Let $G$ be a fully connected, symmetric graph with $N$ vertices.
These two properties mean that every vertex is connected to every
other vertex by an edge and that the same edge is used when travelling
from one vertex to another in either direction.  A vertex is not
connected to itself.  The vertices are labelled $1, 2,\ldots,N$.
$V_i$ is the name for the vertex labelled $i$.  Nothing that follows
depends on the choice of labelling.

Associated with $G$ is the quantum system \qg which is the system
that will simulate the travelling salesman.  In the following,
unless explicitly noted, the graph in question will be the quantum
system \qg and not $G$ because it is \qg that is subject to dynamical
evolution.

I will define a vector space for \qg and assign meaning to these
states.  These states will be sufficient to define a {\it visit\/} of
a vertex and a Hamiltonian tour.  \qg is prepared to be in a special
initial state corresponding to starting at the chosen initial vertex.
A sequence of $N$ operators applied to this state will transform it to
a final state which is a superposition of all Hamiltonian tours and
only the Hamiltonian tours.  Let the initial and final states be $\gstate 0$
and $\gstate N$ respectively and let $\stepop 1,\stepop 2,\ldots,\stepop N$,
be the operators.  The goal is to show that

\begin{equation*}
\gstate N = \stepop N\stepop{N-1}\cdots\stepop 1\gstate 0,
\end{equation*}
and that $\gstate N$ is a superposition of states which may be
interpreted as Hamiltonian tours.  Moreover $\gstate N$ contains only
such states and no others.

\qg is driven from $\gstate 0$ to $\gstate N$ by being coupled to an
external system.  In other words, the step operators $\stepop
1,\stepop 2,\ldots,\stepop N$ are components in a larger Hamiltonian
for the dynamical system consisting a one-dimensional lattice of
spins and \qg \cite{Feynman:1985}.

The vector space for \qg is defined to be the direct product of the
space of each of its vertices.  The space at each vertex is the direct
product of $(N+1)$ qubit spaces.  At each $V_i$,
the basis vectors are

\begin{equation}
\ket{v_i; a_{i1},a_{i2},\ldots,a_{iN}}.\label{vbasis}
\end{equation}

The values of $v_i$, and $a_{ij}$ can be $0$ or $1$.  The basis
vectors span a space of dimension $2^{(N+1)}$ giving \qg a dimension
of $2^{N(N+1)}$.  These states can be transformed by the usual raising
and lowering operators.  So

\begin{displaymath}
\hat v_i\ket{0} = 0,\quad\hat v_i\ket{1} = \ket{0},\quad\hat v_i^*\ket{0} = \ket{1}, \quad\hat v_i^*\ket{1} = 0, \quad i = 1,2,\ldots,N.
\end{displaymath}
Similarly for the states labelled by $a_{ij}$.
\begin{displaymath}
\hat a_{ij}\ket{0} = 0,\quad\hat a_{ij}\ket{1} = \ket{0},\quad\hat a_{ij}^*\ket{0} = \ket{1}, \quad\hat a_{ij}^*\ket{1} = 0, \quad i,j = 1,2,\ldots,N.
\end{displaymath}
These operators obey the well known commutation rules.

\begin{align*}
\hat v_i\hat v_i^* + \hat v_i^*\hat v_i &= 1,\\
\hat v_i\hat v_j^* - \hat v_i^*\hat v_i &= 0, \quad (i \ne j)\\
\hat a_{ij}\hat a_{ij}^* + \hat a_{ij}^*\hat a_{ij} &= 1,\\
\hat a_{ij}\hat a_{kl}^* - \hat a_{ij}^*\hat a_{kl} &= 0, \quad (i,j) \ne (k,l).
\end{align*}
Where $i,j,k,l = 1,2,\ldots,N$.  The $\hat v$ and $\hat v^*$ operators
commute with the $\hat a$ and $\hat a^*$.  The basis vectors of \qg
are

\begin{equation}
\prod_{i=1}^{N}\ket{v_i;a_{i1},a_{i2},\ldots,a_{iN}}.\label{gbasis}
\end{equation}

The meaning of $v_i$, and $a_{ij}$ is as follows.  $\upv i$ means
$V_i$ is being visited while $\downv i$ means $V_i$ is not being
visited.  $\upa i j$ means there is a directed edge from $V_i$ to $V_j$
while $\downa i j$ means there is no such edge.  A state with $\upa i
i$ means there is a self-edge from $V_i$ to itself.  For any $i$ the set
of values $a_{ij}$ represents the edge-state of $V_i$.

\begin{figure}[ht]
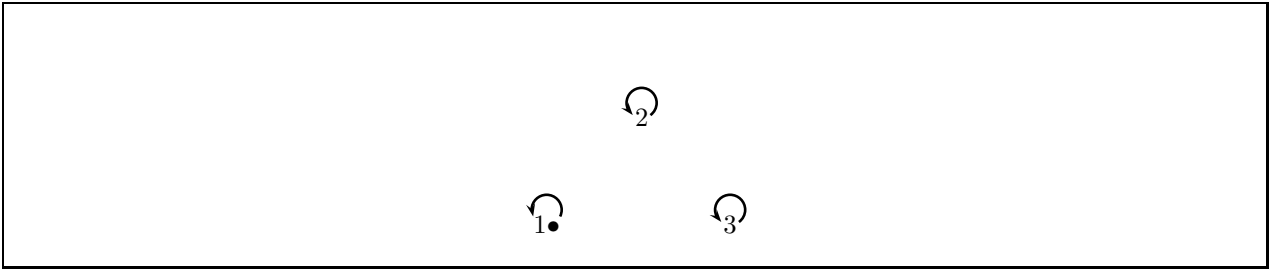

  \begin{center}
$
\psmatrix[colsep=1cm,rowsep=1cm]
\\
&2&\\
1\bullet&&3
\nccircle[linewidth=1pt,nodesep=1pt]{->}{2,2}{.2cm}
\nccircle[linewidth=1pt,nodesep=1pt]{->}{3,1}{.2cm}
\nccircle[linewidth=1pt,nodesep=1pt]{->}{3,3}{.2cm}
\endpsmatrix
$

    \caption{A graph representing the state $\ket {g^{(3)}(0)}$.  The $\bullet$ marks the
      vertex being visited, here $V_1$. Every vertex is connected only to itself with a self edge.}
    \label{v3start}
  \end{center}
\end{figure}

Let us first see how these states can describe a Hamiltonian tour.
Consider a graph $G^{(3)}$, with three vertices $V_1$, $V_2$ and
$V_3$.  The basis vectors of its quantum equivalent $\mathcal G^{(3)}$
are

\begin{equation*}
\ket{v_1;a_{11},a_{12}a_{13}}\ket{v_2;a_{21},a_{22},a_{23}}\ket{v_3;a_{31},a_{32},a_{33}}
\end{equation*}
for all values of the $v_i$ and $a_{ij}$.  Begin by placing its $\mathcal G^{(3)}$ in
state in which one vertex $V_1$, is visited and all three vertices
have only self-edges.  According to the convention, in this state

\begin{align*}
v_1 &= 1, v_2 = v_3 = 0,\\
a_{ij} &\mathrel{\mathop=}
\begin{cases}
1 & \text{if $i=j$;}\\
0 & \text{if $i\ne j$,}
\end{cases}
\end{align*}
where $i,j=1,2,3$.  This is the state

\begin{equation*}
\ket {g^{(3)}(0)} = \ket{1; 1,0,0}\ket{0;0,1,0}\ket{0;0,0,1}.
\end{equation*}
A shorthand for this state is
\begin{equation*}
\ket {g^{(3)}(0)} = \ket{1; \underline{\upa 1 1}}\ket{0; \underline{\upa 2 2}}\ket{0; \underline{\upa 3 3}}.
\end{equation*}
The underline indicates that all $a_{ij}$ not appearing in the ket
have a value of $0$.  In this state, only $v_1=1$ so $V_1$ is the
vertex being visited.  All $a_{ij}=0, (i\ne j)$ and $a_{ii}=1$ so all
vertices have only self-edges.  The graph representing this state is
shown in Figure~\ref{v3start}.

\begin{figure}[ht]
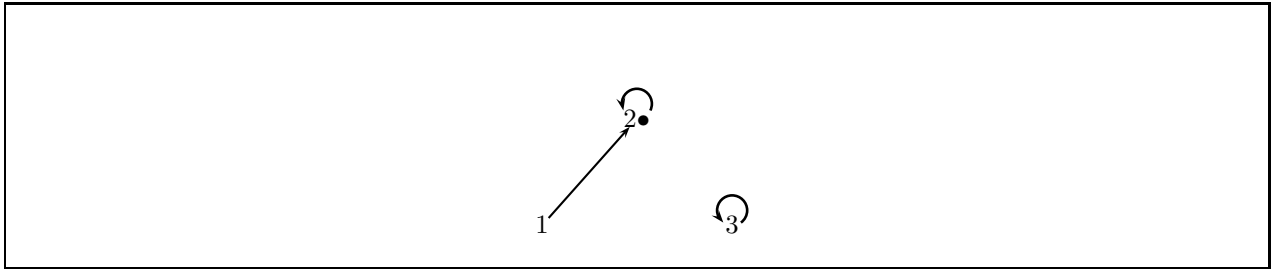

  \begin{center}
$
\psmatrix[colsep=1cm,rowsep=1cm]
\\
&2\bullet&\\
1&&3
\ncline{->}{3,1}{2,2}
\nccircle[linewidth=1pt,nodesep=1pt]{->}{2,2}{.2cm}
\nccircle[linewidth=1pt,nodesep=1pt]{->}{3,3}{.2cm}
\endpsmatrix
$
    \caption{A graph representing the $\ket {0;\underline{\upa 1 2}} \ket {1;\underline{\upa 2 2}} \ket {0; \underline{\upa 3 3}}$
     produced as a result of the first step from $V_1$ to $V_2$.  $V_1$ has no self-edge.  $V_2$ is
     is being visited.}
    \label{v3stepone}
  \end{center}
\end{figure}

Let us take the first step by acting on $\ket {g^{(3)}(0)}$ with the
operator $\down v_1\down a_{11}\up a_{12}\up v_2$.

\begin{equation*}
\down v_1\down a_{11}\up a_{12}\up v_2 \ket {1;\underline{\upa 1 1}}\ket {0;\underline{\upa 2 2}}\ket {0;\underline{\upa 3 3}} \;=\; \ket {0;\underline{\upa 1 2}} \ket {1;\underline{\upa 2 2}} \ket {0; \underline{\upa 3 3}}.
\end{equation*}
In this new state, $v_1\!=\!=0$ and $v_2\!=\!=1$ so by convention this
means $V_2$ is being visited.  The other change is that
$a_{11}\!=\!=0$ while $a_{12}\!=\!=1$ which means there is a directed
edge from $V_1$ to $V_2$.  The effect of the operator $\down v_1\down
a_{11}\up a_{12}\up v_2$ is to make $V_2$ the vertex being visited and
replace the self edge at $V_1$ with a directed edge from $V_1$ to
$V_2$.  The state of $V_3$ is unchanged.  The meaning of this
transformation is that we have carried out the first step of one tour
from $V_1$ to $V_2$.  Figure~\ref{v3stepone} shows the graph
representing this state.

In the second step, visit $V_3$ from $V_2$ with the operator $\down v_2\down a_{22}\up a_{23}\up v_3$.

\begin{equation*}
(\down v_2\down a_{22}\up a_{23}\up v_3)( \down v_1\down a_{11}\up a_{12}\up v_2 ) \ket {g^{(3)}(0)} =\ket {0;\underline{\upa 1 2}} \ket {0;\underline{\upa 2 3}} \ket {1;\underline{\downa 3 3}}
\end{equation*}
After this second step (Figure~\ref{v3steptwo}), $V_3$ is visited and
we have recorded that the visit of $V_3$ took place along the $(2,3)$
edge of $V_2$ ($a_{23}\!=\!1$).  No unvisited vertices remain, so we
must return to the starting vertex with the operator $ \down v_3\down
a_{33}\up a_{31}\up v_1$.  This produces the state
\begin{figure}[ht]
  \begin{center}
$
\psmatrix[colsep=1cm,rowsep=1cm]
&2 \\
1 &&3\bullet
\ncline{->}{2,1}{1,2}
\ncline{->}{1,2}{2,3}
\nccircle[linewidth=1pt,nodesep=1pt]{->}{2,3}{.2cm}
\endpsmatrix
$
    \caption{A graph representing the state $\ket {0;\underline{\upa 1 2}} \ket {0;\underline{\upa 2 3}} \ket {1;\underline{\downa 3 3}}$. $V_3$ is
     is being visited.}
    \label{v3steptwo}
  \end{center}
\end{figure}

\begin{equation*}
( \down v_3\down a_{33}\up a_{31}\up v_1)(\down v_2\down a_{22}\up a_{23}\up v_3)( \down v_1\down a_{11}\up a_{12}\up v_2 )\ket {g^{(3)}(0)}
                 = \ket{1; \underline{\upa 1 2}}\ket{0; \underline{\upa 2 3}}\ket{0; \underline{\upa 3 1}}
\end{equation*}
which is a basis state corresponding to the Hamiltonian tour
$V_1\rightarrow V_2\rightarrow V_3\rightarrow V_1$.  It is carried
out with the three operators $\down v_1\down a_{11}\up a_{12}\up v_2$,
$  \down v_2\down a_{22}\up a_{23}\up v_3$ and $\down v_3\down a_{33}\up a_{31}\up v_1$.

This three-vertex graph has only one other Hamiltonian tour ---
$V_1\rightarrow V_3\rightarrow V_2\rightarrow V_1$ --- which can be
navigated with the operators $\down v_1\down a_{11}\up a_{13}\up
v_3$, $\down v_3\down a_{33}\up a_{32}\up v_2$ and $ \down v_2\down
a_{22}\up a_{21}\up v_1$.  If we can perform both transformations
simultaneously the final state of the graph, ignoring normalisation,
would be

\begin{equation*}
\overbrace{\ket{1;\underline{\upa 1 2}}\ket{0;\underline{\upa 2 3}}\ket{0;\underline{\upa 3 1}}}^{V_1\rightarrow V_2\rightarrow V_3\rightarrow V_1} \;+\;
  \overbrace{\ket{1;\underline{\upa 1 3}}\ket{0;\underline{\upa 3 2}}\ket{0;\underline{\upa 2 1}}}^{V_1\rightarrow V_3\rightarrow V_2\rightarrow V_1}
\end{equation*}
In both terms $V_1$ is the vertex being visited and, since no vertex
has a self edge, all vertices have been visited.  This state is shown
in Figure~\ref{v3final}.
\begin{figure}[ht!]
  \begin{center}
%  \subfloat[$\ket{1; \underline{\upa 1 2}}\ket{0; \underline{\upa 2 3}}\ket{0; \underline{\upa 3 1}}$]
  \subfloat
  {
$
\psmatrix[colsep=1cm,rowsep=1cm]
&2 \\
1 \bullet &&3
\ncline{->}{2,1}{1,2}
\ncline{->}{1,2}{2,3}
\ncline{->}{2,3}{2,1}
\endpsmatrix
$
}

$+$

%  \subfloat[$\ket{1; \underline{\upa 1 2}}\ket{0; \underline{\upa 2 3}}\ket{0; \underline{\upa 3 1}}$]
  \subfloat
  {
$
\psmatrix[colsep=1cm,rowsep=1cm]
&2 \\
1 \bullet &&3
\ncline{->}{2,1}{2,3}
\ncline{->}{2,3}{1,2}
\ncline{->}{1,2}{2,1}
\endpsmatrix
$
}
    \caption{The final state
     $\ket{1; \underline{\upa 1 2}}\ket{0; \underline{\upa 2 3}}\ket{0; \underline{\upa 3 1}} + \ket{1;\underline{\upa 1 3}}\ket{0;\underline{\upa 3 2}}\ket{0;\underline{\upa 2 1}}$.}
    \label{v3final}
  \end{center}
\end{figure}
This example has shown that the basis states in (2) are capable of
encoding a Hamiltonian tour.

A Hamiltonian tour starting at $V_1$ is a basis state with the
following properties ($i,j=1,2,\ldots,N$).
\begin{enumerate}\itemsep0pt
\item $\upv 1$, $\downv i, (i\ne 1)$,
\item $\downa i i$,
\item for each $i$, one $\upa i j$; all other $\downa i j$ ($i\ne j$) and
\item the directed graph represented by the basis state is connected.
\end{enumerate}

In the general case \qg is prepared in an initial state in which
$V_1$ is visited and all vertices are in a state corresponding to
having only self edges.

\begin{equation}
\gstate 0 = \ket{1;\underline{\upa 1 1}}\prod_{j=2}^N\ket{0;\underline{\upa j j}}.\label{gzerodef}
\end{equation}
To take the first step, define the step operator $\hat S(1)$.

\begin{equation}
\hat S(1) = \frac{1}{\sqrt{N-1}}\sum_{i,j=1}^N A_{ij}\down v_{i}\down a_{ii}\up a_{ij}\up v_{j}\label{stepone}
\end{equation}
where $A_{ij}$ is the adjacency matrix of $G$.

\begin{equation*}
A_{ij} =
\begin{cases}
0 & \text{if $i=j$;}\\
1 & \text{if $i\ne j$.}
\end{cases}
\end{equation*}
Let $\gstate 1$ be the state produced by acting on $\gstate 0$ with $\stepop 1$.
\begin{equation*}
\gstate 1 = \stepop 1 \gstate 0 = \frac{1}{\sqrt{N-1}}\sum_{i,j=1}^N A_{ij}\down v_{i}\down a_{ii} \up a_{ij}\up v_{j}\ket{1;\underline{\upa 1 1}}\prod_{k=2}^N\ket{0;\underline{\upa k k}}
\end{equation*}
In the sum over $i$ the only term that survives is the one with
$i\!=\!1$ because $v_1\!=\!1$ and $v_i\!=\!0, (i\!\ne\!1)$. So we have

\begin{equation*}
\gstate 1 = \frac{1}{\sqrt{N-1}}\sum_{j=1}^N A_{1j}\down v_1 \down a_{11}\up a_{1j}\up v_j\ket{1;\underline{\upa 1 1}}\prod_{k=2}^N\ket{0;\underline{\upa k k}}
\end{equation*}
Of the remaining $N$ terms consider the first with $j\!=1\!$ which is
$A_{11}\down v_1\down a_{11}\up a_{11}\up v_1$.  This does not
contribute because $A_{11}\!=\!0$.  Therefore $\gstate 1$ is a
superposition of $(N-1)$ basis states and does not contain a state
resulting from visit of $V_1$ to itself.  After the first step we have

\begin{equation}
\gstate 1 = \stepop 1 \gstate 0 = \frac{1}{\sqrt{N-1}} \sum_{i=2}^N \ket {1\rightarrow i \;\bullet}.\label{gone}
\end{equation}

This is a convenient new notation to represent incomplete tours.
Suppose we want to represent the state for the path $V_1\rightarrow
V_2\rightarrow V_3$.  Here, $V_3$ is being visited from $V_1$ via
$V_2$.  None of the other vertices have been visited and the tour is
currently stopped at $V_3$.  That is, no steps whatsover have been
taken from $V_3$.  We will write this as

\begin{equation*}
\ket {1\rightarrow 2 \rightarrow 3 \; \bullet}.
\end{equation*}
The number appearing last in the sequence identifies the vertex being
visited, in this example $V_3$.  All vertices whose labels appear in the
sequence are in a state that is consistent with this path.  The bullet
after the last number signifies that the remaining vertices, whose
labels do not appear in the sequence, have not taken part in the
tour.  The tour is stopped at the vertex whose number appears last in
the sequence.  Written out fully,

\begin{equation*}
\ket {1\rightarrow 2 \rightarrow 3 \; \bullet} \mathrel{\mathop=^{\rm def}} \ket{0;\underline{a_{12}\!=\!1}}\ket{0;\underline{a_{23}\!=\!1}}\ket{1;\underline{a_{33}\!=\!1}}\prod_{k=4}^{N}\ket{0;\underline{a_{kk}=1}}.
\end{equation*}
In each of the $(N-1)$ terms in equation \eqref{gone} a different
vertex is visited.  $\gstate 1$ is a normalised state by
construction. However, $\hat S(1)$ is not a unitary operator:

\begin{equation*}
\hat S(1)^* \hat S(1) \ne \mathbb{1}
\end{equation*}
But it can be shown that

\begin{align}
\begin{split}
\hat S(1)^* \gstate 1 &= \hat S(1)^* \hat S(1)\gstate 0\\
                      &= \gstate 0.\label{sinverse}
\end{split}
\end{align}
Equation \eqref{sinverse} shows that $\hat S(1)^*$ reverses the action
of $\hat S(1)$ {\it when \qg is initially in the state $\gstate 0$}.
Although $\hat S(1)$ and $\hat S(1)^*$ are not inverses of each other,
they behave as inverses when they act on $\gstate 0$.

For the second step of the tour we define the step operator $\hat S(2)$
which will operate on $\gstate 1$ to produce $\gstate 2$.

\begin{equation*}
\gstate 2 = \hat S(2) \gstate 1,
\end{equation*}
where
\begin{equation*}
\hat S(2) = \frac{1}{\sqrt{N-2}}\sum_{i,j=1}^N A_{ij}\down v_i\down a_{ii}\up a_{ij}\up a_{jj}\down a_{jj}\up v_j.
\end{equation*}
$\hat S(2)$ is identical to $\hat S(1)$ except that it contains a new
operator combination $\up a_{jj}\down a_{jj}$ and a normalisation
factor of ${(N-2)}^{-1/2}$ not ${(N-1)}^{-1/2}$.  The role of the
additional operator combination is described presently. $\hat S(2)$
will act on the $(N-1)$ terms of $\gstate 1$ in equation \eqref{gone}
to produce many more terms.  Let us look at $\hat S(2)$ acting on just
the first term of $\gstate 1$, namely $\ket{1\rightarrow 2\;\bullet}$.
In the sum over $i$ the only term that survives is the one with
$i\!=\!2$ as all others are annihilated by the action of $\hat
v_i$. From $N^2$ terms we are reduced to $N$.

\begin{equation*}
\hat S(2) \ket {1\rightarrow 2 \;\bullet} = \frac{1}{\sqrt{N-2}}\sum_{j=1}^N A_{2j}\down v_2\down a_{22}\up a_{2j}\up a_{jj}\down a_{jj}\up v_j\ket {1\rightarrow 2 \;\bullet}.
\end{equation*}
The term with $j\!=\!2$ does not contribute because $A_{22}\!=\!0$ so
the sum over $j$ produces at most $(N-1)$ terms.  The term with
$j\!=\!1$ corresponds to visiting $V_1$ from $V_2$ which would
certainly not be a legal step since $V_1$, being the starting vertex,
has already been visited.  But the operator for this step contains
$\up a_{11}\down a_{11}$ which will produce $0$ because in this state
$a_{11}\!=\!0$.  So the term with $j\!=\!1$ does not survive meaning
that $V_1$ is not visited from $V_2$ in the second step. So the sum
over $j$ produces just $(N\!-\!2)$ terms.  Since there is nothing
special about $V_2$, the same argument holds for all other vertices.
Therefore, from each term of $\gstate 1$ in equation \eqref{gone}, the
action of $\hat S(2)$ will produce $(N\!-\!2)$ terms to give

\begin{equation*}
\gstate 2 = \hat S(2) \hat S(1) \gstate 0 = \frac{1}{\sqrt{(N\!-\!1)(N\!-\!2)}}\sum_{i\!\ne\!1}^N\sum_{\substack{j\!\ne\!1\\ j\!\ne\!i}}^N\ket {1\rightarrow i\rightarrow j\;\bullet}.
\end{equation*}
$\gstate 2$ is normalised.  The operator $\hat S(2)^*$ reverts $\gstate 2$
to $\gstate 1$.

\begin{align*}
\hat S(2)^* \hat S(2) \gstate 2 &= \frac{1}{\sqrt{N\!-\!1}}\frac{1}{\sqrt{N\!-\!2}} \sum_{i=2}^N(N-2) \ket{1\rightarrow i\;\bullet}\\
                                &= \frac{1}{\sqrt{N\!-\!1}}\sum_{i=2}^N \ket{1\rightarrow i\;\bullet}.
\end{align*}
It is convenient then, to define the adjacency operator $\hat A_{ij}(T)$

\begin{equation}
\hat A_{ij}(T) = \frac{1}{\sqrt{N-T}} A_{ij}\down a_{ii}\up a_{ij}\up a_{jj}\down a_{jj}, \quad T=1,2,\ldots,(N\!-\!1)\label{aij}
\end{equation}
where $i=1,2,\ldots,N$, and from which we can define a sequence of step
operators, $\hat S(T)$, for the $N$ steps of the tour.

\begin{equation}
\hat S(T) =
\begin{cases}
\sum_{i,j=1}^N\down v_i\hat A_{ij}(T) \down v_j, & \text{if $T=1,2,\ldots,(N-1)$}\\
\noalign{\vskip 8pt}
\sum_{i=1}^N \down v_i \down a_{ii}\up a_{i1}\up v_1, & \text{if $T=N$}. \label{allops}
\end{cases}
\end{equation}
The operator for the last step, $\hat S(N)$, does not produce new
terms.  Its effect is to visit $V_1$ for the second time.  After the
action of $\hat S(N)$ every term in the superposition will have
$v_1\!=\!1$.  Starting from the initial state $\gstate 0$ we produce a
final state $\gstate N$ where

\begin{equation}
\gstate N = \hat S(N)\hat S(N-1)\cdots\hat S(1) \gstate 0.\label{gn}
\end{equation}
$\gstate N$ is a superposition of $(N-1)!$ basis states and is
normalised by construction.  Each of these corresponds to a distinct
Hamiltonian tour.  All terms in the superposition have the same
probability amplitude of ${1/\sqrt{(N-1)!}}$.

As has been noted, $\hat S(T)$ is not a unitary operator.  Although the
operators used in Feynman's description are unitary, the following weaker
condition is sufficient.

\begin{equation*}
\hat S(T)^*\hat S(T)\gstate {T-1} = \gstate {T-1}
\end{equation*}
This implies that the computation is meaningful only starting from the
special state $\gstate 0$.

\section{The cost of a tour}

In the simulation just described all tours have the same quantum
amplitude.  If a measurement is made on $\gstate N$, one tour is as
likely as another.  This can be improved by taking into account the
cost of a tour.  Let $w_{ij}$ be the weight of the edge $(i, j)$
defined as follows.

\begin{equation}
w_{ij} = 0 \quad \text{if $i=j$}; \qquad 0 < w_{ij} < 1 \quad \text{if $i\ne j$.}
\end{equation}
The $w_{ij}$ are non-negative real numbers which are normalised.

\begin{equation}
\sum_{j=1}^N w_{ij}^2 =1, \qquad(i=1,2,\ldots,N).\label{normw}
\end{equation}
The $w_{ij}$ are taken to be symmetric: $w_{ij} = w_{ji}$.  We can
always impose on the $w_{ij}$ the $N$ conditions in equation
\eqref{normw} because there are $N(N-1)/2$ $w_{ij}$.  In equation
\eqref{aij} we defined $\hat A_{ij}(1)$ the operator for the first
step

\begin{equation*}
\hat A_{ij}(1) = \frac{1}{\sqrt{N-1}} A_{ij}\down a_{ii}\up a_{ij}\up a_{jj} \down a_{jj}, \quad i,j=1,2,\dots,N.
\end{equation*}
We replace this with a new definition
\begin{equation}
\hat A_{ij}(1) = w_{ij}\down a_{ii}\up a_{ij}\up a_{jj} \down a_{jj}.\label{newstepone}
\end{equation}
For the first step the operator $\hat S(1)$ will use this new $\hat
A_{ij}(1)$ rather than the one defined in equation \eqref{aij}.
Acting on $\gstate 0$ with the new $\hat A_{ij}(1)$ gives

\begin{equation*}
\gstate 1 = \hat S_1 \gstate 0 = \sum_{i=2}^N w_{1i} \ket{1\rightarrow i\; \bullet}.
\end{equation*}
Because of the condition in equation \eqref{normw}, $\gstate 1$ is
normalised and it is still the case that

\begin{equation*}
\hat S_1^* \gstate 1 = \hat S_1^* \hat S_1 \gstate 0 = \gstate 0.
\end{equation*}

The amplitude for the step $V_1\rightarrow V_i$, which was previously
${(N-1)}^{-1/2}$, is now $w_{1i}$.  For the second step, suppose we
could define an operator that is identical to $\hat A_{ij}(1)$ in
equation \eqref{newstepone}.

\begin{equation*}
\hat A_{ij}(2) = \hat A_{ij}(1)
\end{equation*}
and operate on $\gstate 1$.  As before $\gstate 2$ will contain
$(N-1)(N-2)$ terms because we have changed only the amplitudes and not
the dynamics.  After the second step the path $V_1\rightarrow
V_{i_1}\rightarrow V_{i_2}$ will have quantum amplitude
$w_{1i_1}w_{i_1i_2}$.  But recall that in the second step, starting
from $V_{i_1}$, we do not visit $V_1$ or $V_{i_1}$ itself.  That is,
only $(N-2)$ other vertices are visited not $(N-1)$.  In any step,
starting from a given vertex, the total probablity of all terms
produced must be unity.  The total probability of all terms produced by
travelling from $V_{i_1}$ to $(N-2)$ other vertices is produced by
summing over $i_2$ --- the index of the vertices visited from
$V_{i_1}$.

\begin{equation*}
\sum_{{i_2}\ne 1}^Nw_{{i_1}{i_2}}^2 \;=\; (1 - w_{{i_1}1}^2) \;<\; 1.
\end{equation*}
Therefore $\gstate 2 = \hat S(2)\gstate 1$ is not a normalised state.
We are missing a probability of $w_{{i_1}1}^2$.  We can get around this
situation by using a modified operator for the second step

\begin{equation}
\hat A_{ij}(2) = \frac{w_{ij}}{\sqrt{1 - w_{i1}^2}}\;\down a_{ii}\up a_{ij}\up a_{jj} \down a_{jj}.\label{newsteptwo}
\end{equation}

This is exactly the same as $\hat A_{ij}(1)$ in equation \eqref{newstepone}
except that all weights have been rescaled.  In terms of the
rescaled weights, in the second step, the total probability of all
terms produced by starting from $V_{i_1}$ is now

\begin{align*}
\sum_{{i_2}\ne 1}^N \frac{w_{{i_1}{i_2}}^2}{(1- w_{{i_1}1}^2)} &=  \frac{1}{(1- w_{{i_1}1}^2)} \sum_{{i_2}\ne 1}^N w_{{i_1}{i_2}}^2\\
                                                     &= 1
\end{align*}
as required.  The probability of all paths starting from
$V_1$ is produced by summing over ${i_1}$ and ${i_2}$.

\begin{align*}
\sum_{i_1, i_2, {i_2}\ne 1}^N w_{1i_1}^2\frac{w_{{i_1}{i_2}}^2}{(1- w_{{i_1}1}^2)} &=  \sum_{i_1=1}^N\frac{w_{1i_1}^2}{(1- w_{{i_1}1}^2)} \sum_{{i_2}\ne 1}^N w_{{i_1}{i_2}}^2\\
                                                     &= \sum_{i_1=1}^Nw_{1i_1}^2.
\end{align*}
This produces, from equation \eqref{normw}, unity proving that $\gstate 2$ is
normalised.

The rescaling of weights works in step $2$ because in this step, every
vertex, $V_{i_1}$, visited in the first step was visited from the
fixed vertex $V_1$.  At the end of the first step, every path has the
same history --- all start at $V_1$.  This is not the case in
subsequent steps.  In the third step look for example, at the path
$V_1\rightarrow V_2\rightarrow V_3 \rightarrow V_4$.  In going from
$V_3$ to $V_4$, neither $V_1$ nor $V_2$ is visited from $V_3$, the
missing probablity is $(w_{31}^2 + w_{32}^2)$ which means all
amplitudes for the step starting from $V_3$ must be scaled by $(1 -
(w_{31}^2 + w_{32}^2))^{-1/2}$.  Now look at the path $V_1\rightarrow
V_3\rightarrow V_4 \rightarrow V_6$.  Here the missing probability in
going from $V_4$ to $V_6$ is $(w_{41}^2 + w_{43}^2)$.  So for this
path the scaling factor is $(1 - (w_{41}^2 + w_{43}^2))^{-1/2}$.  For
the second and subsequent steps every path has a distinct history
which cannot be accounted for by the technique used in step 2.

We need an operator to rescale the weights.  Let us look at this one.

\begin{equation}
\hat{w_i} = \sum_{j=1}^N w_{ij}^2\down a_{jj} \up a_{jj}.\label{wi}
\end{equation}
This is a real, linear operator which has every basis state of $G$ as
an eigenvector.  Let us apply $\hat{w_3}$ to the path $V_1\rightarrow
V_2 \rightarrow V_3$.

\begin{equation*}
\hat w_3\ket{1\rightarrow 2\rightarrow 3\;\bullet} = (w_{31}^2 + w_{32}^2) \ket{1\rightarrow 2\rightarrow 3\;\bullet}.
\end{equation*}
So $\hat{w_3}$ picks out the total probability along all edges from
$V_3$ leading vertices already visited along the path $V_1\rightarrow
V_2 \rightarrow V_3$.  In general, if along some path, $V_i$ is the
vertex being visited and we operate on that state with $w_i$, the
eigenvalue will be the sum of the squares of weights from $V_i$ to all
vertices which have already been visited along the path.  However the
operator we really want is one related to $\hat w_i$.  Look at the
rescaling operator

\begin{equation*}
\widehat W_i = \Bigl(1 - \sum_{j=1}^N w_{ij}^2\up a_{jj} \down a_{jj}\Bigr)^{-1/2}
\end{equation*}
which we can write as a power series

\begin{equation*}
\widehat W_i = c_0 + c_1 \sum_{j=1}^N w_{ij}^2\up a_{jj} \down a_{jj} + c_2 \Bigl(\sum_{j=1}^N w_{ij}^2\up a_{jj} \down a_{jj}\Bigr)^2 + \cdots.
\end{equation*}
where the $(c_0, c_1, \ldots)$ are real numbers.  This is a power
series of commuting operators $\up a_{ii}\down a_{ii}$.  So ${\widehat
  W_i}$ is a linear operator and can be used to renormalize weights along
the edges.  The problem is that the $\widehat W_i$ are highly
degenerate and have a large number of eigenvalues which are
infinite. Any state corresponding to a Hamiltonian tour --- in which
all vertices have been visited --- will be a state with an infinite
eigenvalue.  A Hamiltonian tour is a state in which $a_{jj}\!=\!0$ for
every $j$.  In this state, for any fixed $i$ and distinct
$i_1,i_2,\ldots,i_{(N-1)}$,

\begin{align*}
(1 - \sum_{j=1}^N w_{ij}^2\down a_{jj} \up a_{jj})\ket{1\rightarrow i_1\rightarrow \cdots \rightarrow i_{(N-1)}\rightarrow 1} &= (1 -\sum_{j=1}^N w_{ij}^2)\ket{1\rightarrow i_1\rightarrow \cdots \rightarrow i_{(N-1)} \rightarrow 1}\\
&= 0.
\end{align*}
To avoid this we redefine the scaling operator.

\begin{equation}
\widehat W_i = \Bigl(1 - \hat v_1\hat v_1^*\sum_{j=1}^N w_{ij}^2\up a_{jj} \down a_{jj}\Bigr)^{-1/2}.\label{mhat}
\end{equation}
The infinite eigenvalues are avoided because the action of $\hat
v_1\hat v_1^*$ in equation \eqref{mhat} will produce $0$ for any state in which
$v_1\!=\!1$.  At both the start and end of a tour the second term
in equation \eqref{mhat} vanishes to produce a scaling of unity which is
exactly what is required.

The adjacency operator for the graph is now

\begin{equation*}
\hat A_{ij} =  w_{ij}\down a_{ii}\up a_{ij}\up a_{jj} \down a_{jj} \widehat W_i.
\end{equation*}
The step operators are

\begin{equation}
\hat S(T) =
\begin{cases}
\sum_{i,j=1}^N \down v_i\hat A_{ij}(T)\up v_j , & \text{if $T=1,2,\ldots(N-1)$}\\
\noalign{\vskip 8pt}
\sum_{i=2}^N\down v_i \down a_{ii}\up a_{i1}\up v_1, & \text{if $T=N$}.
\end{cases}
\end{equation}
These will produce all Hamiltonian tours as desired.  The amplitude
for the last step is always unity.  This is because in the last step
we must return to the starting vertex.  But in fact it is unity even
in step $(N-1)$.  Scaling increases the weights along the edges which
is expected since after each step there are fewer edges along which to
leave a given vertex.  The amplitude for the tour
$V_1\rightarrow V_{i_1}\rightarrow V_{i_2}\rightarrow V_{i_3}\rightarrow\cdots\rightarrow V_{i_N}\rightarrow V_1$ will be

\begin{equation}
w_{1i_1}\frac{w_{i_1i_2}}{\sqrt{1 - w_{i_11}^2}}\frac{w_{i_2i_3}}{\sqrt{1 - (w_{i_21}^2 + w_{i_2i_1}^2)}}\cdots 1.\label{amp}
\end{equation}
where each term in the product is the factor produced in each step.  Each
factor is a ratio that approaches unity as the tour progresses.

To connect with the classical result, let $c_{ij}$ be the classical
cost of travelling along the edge $(i, j)$.  Each $c_{ij}$ is a
positive real number with $c_{ij} = c_{ji}$ and $c_{ii}\!=\!0$.  Let

\begin{equation}
(C_i)^2 = \sum_{j=1, j\ne i}^Ne^{-2c_{ij}}
\end{equation}
An obvious choice for $w_{ij}$ is

\begin{equation}
w_{ij} = 
\begin{cases}
0, & \text{if $i=j$;}\\
\frac{e^{-c_{ij}}}{C_i}, & \text{if $i\ne j$.}
\end{cases}
\end{equation}
These $w_{ij}$ satisfy the condition in equation \eqref{normw} and can be
treated as quantum amplitudes.  Smaller classical costs $c_{ij}$ will produce
larger quantum amplitudes for a step along the corresponding edge.

In terms of the $c_{ij}$ equation \eqref{amp} is

\begin{equation*}
\frac{e^{-c_{1i_1}}}{C_1}\,\frac{e^{-c_{i_1i_2}}}{\sqrt{{C_{i_1}}^2 - e^{-2c_{i_11}}}}\,\frac{e^{-c_{i_2i_3}}}{\sqrt{{C_{i_2}}^2 - (e^{-2C_{i_21}} + e^{-2c_{i_2i_1}})}}\cdots \frac{e^{-c_{i_N1}}}{e^{-c_{i_N1}}}.
%% + c_{ij} + c_{jk} + \cdots)}}{C_1\sqrt{{C_i}^2 - e^{-2c_{i1}}}\sqrt{{C_j}^2 - e^{-(2c_{j1} + 2c_{ji})}}\cdots}
\end{equation*}
But for scaling, which is an effect of the simulation, the quantum
amplitude is the exponential of the sum of the classical costs of the
edges in the tour.

The scaling factors have two quite different effects.  In the
classical analysis, the Hamiltonian tour $V_1\rightarrow
V_2\rightarrow V_3\cdots\rightarrow V_N\rightarrow V_1$ has the same
cost as the tour in the reverse direction.  In this quantum simulation
these two tours will generally have different quantum amplitudes
because the scaling factors will be different for each.

In the classical case the optimal paths are those with the smallest
total cost.  In this simulation the optimal paths are those with the
largest quantum amplitude.  Due to scaling, Hamiltonian tours with the
largest quantum amplitudes may not be the same as the classical tours
with the smallest classical cost.  The two results are expected to be
the same in the majority of cases but this is based on the general
notion that the classical optimal path will play the leading role even
in the quantum domain.  Numerical simulations on small problems and
a few choices of edge-costs seem to support this claim.

\section{Summary}

By doing quantum mechanics on a graph I have described a simulation
which can efficiently traverse all Hamiltonian tours and only the
Hamiltonian tours.  It is expected that in the majority of cases,
tours with the smallest classical cost will have the largest quantum
amplitude.  There will certainly be cases in which the two differ.
One can take this to mean that this simulation does not always produce
a final superposition containing the correct --- classically defined
--- results.  On the other hand the final state contains basis states
which represent the quantum optimal paths.  Computing either is
equally difficult.

It will take $O(N(N-1)/2)$ measurements on the final state to reduce
it to a basis state corresponding to a Hamiltonian tour.  In general
the amplitude for any give tour is $O({1/\sqrt{(N-1)!}}$ so as
expected, the simulation will have to be executed many times to find
the optimal paths.  This simulation will therefore be effective in
those cases where the optimal tours have much larger quantum
amplitudes than the sub-optimal tours.

The scaling operators $\widehat W_i$ have additional infinities.
States with $v_1=0$ and all $a_{ii}=0$ are states with an infinite
eigenvalue.  However these states are never reached in this
simulation.

The marking scheme described in section 2 is by no means unique.
There is plenty of room for different choices of basis states and the
step operators.  The one described here is natural because of the
central role of the adjacency matrix.  It is also one of the simplest.
The marking scheme described in section 2 is an efficient way to
traverse all Hamiltonian tours without any scaling.

One of the objectives of this simulation was to make quantum
amplitudes depend in some way on the classical cost of travelling
along an edge.  Other simulations are possible in which the classical
costs are used in different ways.

\section{Acknowlegements}
I would like to thank Wade Walker for his encouragement and careful
reading of the manuscript.


\begin{thebibliography}{1}

\bibitem{tspbook:1985}
E.~L. Lawler, J.~K. Lenstra, A.~H.~G. Rinnooy~Kan, and D.~G. Sohmoys.
\newblock {\em The {T}raveling {S}alesman {P}roblem: A {G}uided {T}our of
  {C}ombinatorial {O}ptimization}.
\newblock Wiley, 1985.

\bibitem{Deutsch:1985}
D.~Deutsch.
\newblock Quantum theory, the {C}hurch-{T}uring principle and the univeral
  quantum computer.
\newblock {\em Proceedings of the Royal Society of London}, A 400:97--117,
  1985.

\bibitem{Benioff:1982}
P.~Benioff.
\newblock Quantum {M}echanical {M}odels of {T}uring {M}achines that {D}issipate
  {N}o {E}nergy.
\newblock {\em Phys. Rev. Lett.}, 48:1581--1585, 1982.

\bibitem{Feynman:1985}
R.~P. Feynman.
\newblock Quantum {M}echanical {C}omputers.
\newblock {\em Optics News}, 11(2):11--20, 1985.

\bibitem{Lloyd:1996}
S.~Lloyd.
\newblock Universal {Q}uantum {S}imulators.
\newblock {\em Science}, 273(5278):1073--1078, 1996.

\end{thebibliography}
\end{document}